\documentclass[useAMS,usenatbib,usegraphicx]{mn2e}

\title[Abundances in HD\,27411]{Abundances in HD\,27411 and the helium
problem in Am stars}
\author[G. Catanzaro, L. A. Balona]{G. Catanzaro$^{1}$\thanks{E-mail: gca@oact.inaf.it}, 
L. A. Balona$^{2}$\\
$^{1}$INAF - Osservatorio Astrofisico di Catania, Via S. Sofia 78, I--95123, Catania, Italy\\
$^{2}$South African Astronomical Observatory, P.O. Box 9, Observatory 7935, Cape Town, South Africa\\
}
\begin{document}

\date{Accepted . Received ; in original form }

\pagerange{\pageref{firstpage}--\pageref{lastpage}} \pubyear{2002}

\maketitle

\label{firstpage}

\begin{abstract}
We analyze a high-resolution spectrum of the A3m star HD\,27411.   We compare 
abundances derived from ATLAS9 model atmospheres with those using the more 
computationally-intensive ATLAS12 code.  We found very little
differences in the abundances, suggesting that ATLAS9 can be used for
moderate chemical peculiarity.  Our abundances agree well with the
predictions of diffusion theory, though for some elements it was necessary
to calculate line profiles in non-thermodynamic equilibrium to obtain agreement.
We investigate the effective temperatures and luminosities
of Am/Fm stars using synthetic Str\"{o}mgren indices derived from calculated 
spectra with the atmospheric abundances of HD\,27411.  We find that the 
effective temperatures of Am/Fm stars derived from Str\"{o}mgren photometry are
reliable, but the luminosities are probably too low.  Caution is required
when deriving the reddening of these stars owing to line blanketing effects.
A comparison of the relative proportions of pulsating and non-pulsating Am
stars with $\delta$~Scuti stars shows quite clearly that there is no
significant decrease of helium in the driving zone, contrary to current
models of diffusion.
\end{abstract}


\section{Introduction}

The ``metallic-lined'' or Am stars are A-type stars which have strong absorption 
lines of some metals such as Zn, Sr, Zr and Ba and weaker lines of other metals 
such as Ca and/or Sc relative to their spectral type as determined by the strength 
of the hydrogen lines \citep{preston74}.  The strong metallic lines are 
more typical of an F star rather than an A star.  The work of
\citet{michaud70} established radiative diffusion in a strong magnetic field as 
the likely cause of the chemical peculiarities in Ap stars.  When the
magnetic field is absent, diffusion leads to the Am/Fm stars \citep{watson71}.
  The presence of magnetic fields in Am stars has been investigated, but 
with negative results, (e.g. \citet{fossati07}). A peculiarity of Am stars is that 
their projected rotational velocities are generally much smaller than normal A 
stars and they are nearly always members of close binary systems.  Rotational 
braking by tidal friction in a binary system is regarded as a possible 
explanation for the low rotational velocities in Am stars. Slow rotation 
further assists the segregation of elements by diffusion.

The abundance anomalies predicted by the diffusion hypothesis are usually 
much larger than observed.  \citet{richer00} developed detailed models of the 
structure and evolution of Am/Fm stars using OPAL opacities, taking into 
account atomic diffusion and the effect of radiative acceleration.  These 
models develop a convective zone due to ionization of iron-group elements at 
a temperature of approximately 200,000~K.  In addition to this convective zone, 
these stars also have a thin superficial convective zone in which H and He{\sc i} 
are partially ionized.  By assuming sufficient overshoot due to turbulence, 
these separate convective zones become one large convective zone.  The
resulting mixing dilutes the large abundance anomalies predicted by previous 
model, leading to abundances which closely resemble those observed in Am/Fm stars.  

A detailed abundance analysis of eight Am stars belonging to the Praesepe 
cluster \citep{fossati07} show good agreement with the predictions of 
\citet{richer00} for almost all the common elements except for Na and 
possibly S.  The models of \citet{richer00} assume a certain ad-hoc 
parametrization of turbulent transport coefficients which are adjusted to 
reproduce observations.  Other parameterizations of turbulence have been 
proposed for other types of stars.  \cite{talon06} have investigated to what 
extent these are consistent with the anomalies observed on Am/Fm stars.  They 
find that the precision of current abundances is insufficient to distinguish 
between models.  More recently, \citet{michaud11} have studied the
abundance anomalies of the mild Am star Sirius A.  They find that except for
B, N and Na, there is good agreement with the predicted anomalies but
turbulent mixing or mass loss is required.  It is not clear whether it is
turbulence or mass loss which competes with diffusion to lower the abundance
anomalies.  For example, \citet{vick11} find that diffusion in the presence of
weak mass loss can explain the observed abundance anomalies of
pre-main-sequence stars.  This is in contrast to turbulence models which do 
not allow for abundance anomalies to develop on the pre-main-sequence.

Most of the pulsational driving in $\delta$~Scuti stars is caused by the 
$\kappa$~mechanism operating in the He{\sc ii} ionization zone.  Diffusion tends
to drain He from this zone and therefore pulsational driving may be expected
to be weaker or absent in Am/Fm stars \citep{baglin72}. In fact, for many years 
it was thought that classical Am/Fm stars did not pulsate, though claims were
made for some stars \citep{kurtz89}.  Recently, intensive ground-based observations
by SUPER-WASP \citep{smalley11}, and also from  the {\it Kepler} mission 
\citep{balona11} have shown that many Am/Fm stars do pulsate.  \citet{smalley11},
for example, found that about 200 Am/Fm stars out of a total of 1600 (12.5
percent) show $\delta$~Sct pulsations, but with generally lower amplitudes.  
They found that the pulsating Am/Fm stars are confined between the red and
blue radial fundamental edges, in agreement with \citet{balona11}.  While
there are many $\delta$~Sct stars hotter than the fundamental blue edge,
this does not seem to be the case for pulsating Am/Fm stars.  The
significance of this result remains to be evaluated. 

The effect of draining of He from the He{\sc ii} ionization zone is to
reduce the width of the instability strip, the blue edge moving towards the
red edge, eventually leading to the disappearance of the instability strip
when He is sufficiently depleted \citep{cox79}.  \citet{turcotte00} has 
discussed the effect of diffusion on pulsations in Am/Fm stars using the 
models by \citet{richer00}.  One significant difference with earlier models 
is that a substantial amount of He remains in the He{\sc ii} ionization zone.
The blue edge of the instability strip for Am/Fm stars is sensitive to the 
magnitude of the abundance variations and is thus indicative of the depth
of mixing by turbulence.  \citet{turcotte00} predict that pulsating Am/Fm 
stars should lie in a confined region of the HR diagram close to the red edge 
of the $\delta$~Sct instability strip.  However, \citet{balona11} show that 
there is no relationship between the predicted Am/Fm instability strip and the 
actual location of these stars in the HR diagram.

A particularly interesting result of the pulsation analysis of \citet{turcotte00} 
is the prediction of long-period g modes in A-type stars.  As the star
evolves, the driving regions shift deeper into the star and the g modes 
become gradually more and more excited.  Whereas p modes are stabilized 
through diffusion, g modes tend to be excited as a result of that process. 
It appears that diffusion may act to enhance driving of long-period g modes
due to a significant increase in opacity due to iron-group elements.
This may have a bearing on the fact that nearly all A-type stars observed by
{\it Kepler} have unexplained low-frequencies \citep{balona11a}.

When dealing with objects with non-standard chemical composition, such as Am 
stars, it is crucial that the opacities are correctly calculated.  This question 
has been investigated by several authors in recent years.  These studies 
show that a non-standard chemical composition of the stellar atmosphere alters 
the flux distribution of the star or modifies the profiles of the Balmer lines 
(\citet{leo97}, \citet{cat04}). Therefore a determination of T$_{\rm eff}$ and 
$\log g$ based on a comparison between observed and computed Balmer-line profiles 
will not be correct unless one takes into account the metallicity of the star.   
Thus, even estimates based on standard analysis of the spectra may be in error
when applied to Am/Fm stars.

In this paper, we investigate the determination of effective temperature and
surface gravity of the Am star HD\,27411 (HR~1353, A3m) using spectra in the ESO
archives.  The purpose is to determine whether the stellar parameters of
this star agree with those obtained from Str\"{o}mgren photometry and hence
to test the reliability of the effective temperature calibration applied to
Am/Fm stars.  The star was used by \citet{ryab08} as a comparison in their study on 
the calcium stratification in Ap stars.  HD\,27411 is not known to pulsate.
However, as we know from {\it Kepler} observations, pulsations in A and F stars 
with amplitudes too low to be detected from the ground are common. 

Atmospheric models obtained with ATLAS9 \citep{kur93} use precomputed line opacities 
in the form of opacity distribution functions (ODFs).  These are tabulated for 
multiples of the solar metallicity and for various microturbulent velocities.  This 
approach allows very fast computation of model atmospheres, but with very
little flexibility in choice of chemical profile and microturbulent velocity. 
While this is satisfactory for most applications, it fails for chemically
peculiar stars where a non-standard chemical composition profile is
required.  This can be done with ATLAS12 \citep{kur97}, which is essentially 
identical to ATLAS9, but uses the opacity sampling (OS) method to evaluate line 
opacities.  In this study we compare the abundances of HD\,27411 obtained with 
both codes to determine if the use of ATLAS12 is essential.

\begin{figure*}
\centering
\includegraphics[width=8.8cm]{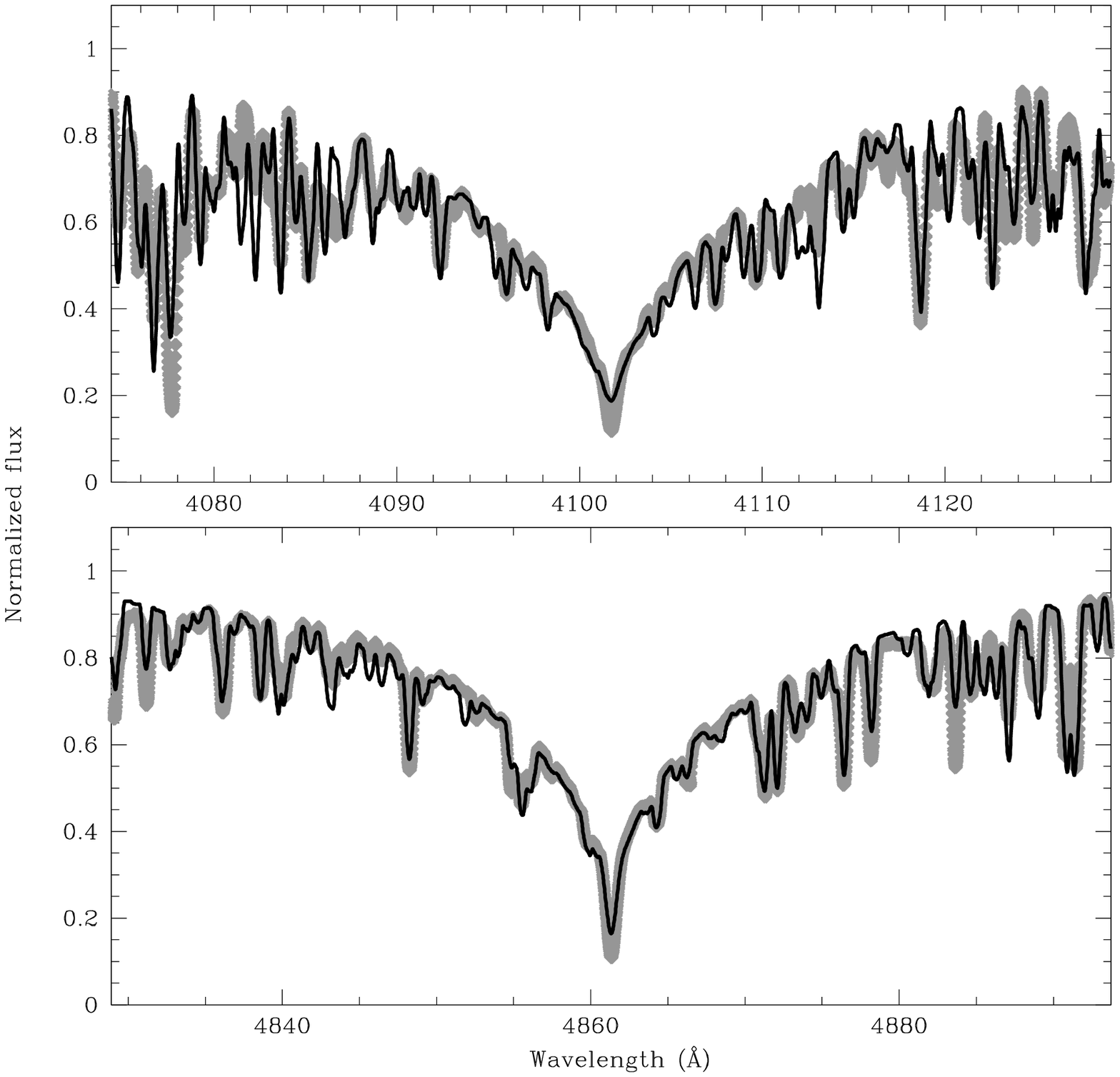}
\includegraphics[width=8.8cm]{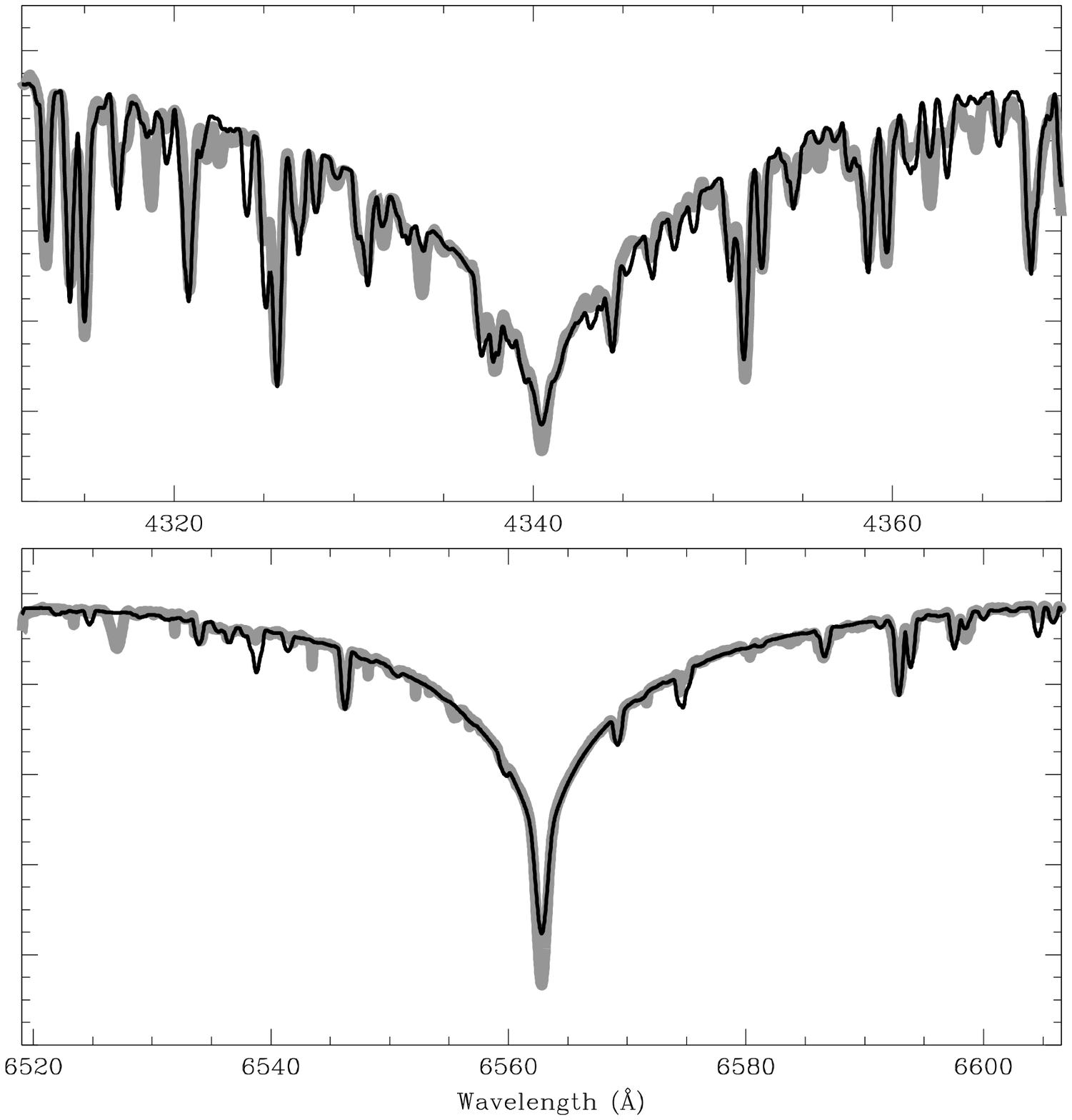}
\caption{Comparison between the observed (crosses) and computed (solid red line) hydrogen line profiles. From top 
to bottom: the Balmer line profiles from H$_{\delta}$ to H$_{\alpha}$. The synthetic 
profiles were computed with SYNTHE using an ATLAS12 model atmosphere with 
T$_{\rm eff}$\,=\,7400~$\pm$~150~K, $\log g$\,=\,4.0~$\pm$~0.1, $\xi$\,=\,4.2~$\pm$~0.3~km~s$^{-1}$, 
$v_e \sin i$\,=\,20.5\,$\pm$\,0.5~km~s$^{-1}$ and individual abundances
shown in Table~\ref{abund}.}
\label{balmer}
\end{figure*}

The result that Am stars are not confined to particular region of the
$\delta$~Sct instability strip depends, to a large extent, on effective 
temperatures and luminosities estimated from  Str\"{o}mgren photometry
\citep{smalley11}.  It is not clear whether the calibration, derived from 
normal AF stars, can be applied to Am/Fm stars.  In this paper we use 
synthetic Str\"{o}mgren photometry applied to models of Am/Fm stars to 
investigate the reliability of fundamental parameters estimated from the
photometry.  Finally, we discuss the relative numbers of pulsating and
non-pulsating Am stars and compare these to the relative numbers of
$\delta$~Scuti and constant stars in the instability strip.  From this
comparison, one can deduce the effectiveness of pulsational driving in the
He{\sc ii} ionization zone and compare the He abundance to that expected
from diffusion calculations.

 \begin{table}
 \centering
  \caption{Ions used to determine the microturbulent velocity in HD\,27411.  The 
number of spectral lines used, the microturbulent velocity, $\xi$, the derived 
abundance and the radial velocity, RV, are listed.}
  \begin{tabular}{lcccc}
  \hline
  \hline
  Elem & N & $\xi$ & Abundance & RV \\
       &   & km s$^{-1}$ & $\log N_{\rm el}/N_{\rm Tot}$ & km s$^{-1}$ \\
 \hline
 Fe{\sc i}  & 71 & 3.9\,$\pm$\,0.3 & $-$3.80\,$\pm$\,0.03 & 40.4\,$\pm$\,0.6\\
 Fe{\sc ii} & 15 & 4.4\,$\pm$\,0.2 & $-$3.84\,$\pm$\,0.04 & 40.9\,$\pm$\,0.6\\
 Ni{\sc i}  & 24 & 4.4\,$\pm$\,0.6 & $-$4.63\,$\pm$\,0.03 & 40.5\,$\pm$\,1.3\\
 \hline
 Adopted   & --  & 4.2\,$\pm$\,0.3 & -- & 40.6\,$\pm$\,0.3\\
 \hline
\end{tabular}
\label{xi_vr}
\end{table}

\section{Observation and data analysis}

The spectrum of HD\,27411 is available in the ESO archive as a part of the 
UVES Paranal Observatory Project (UVES POP), which aims to create a library 
of high-resolution spectra across the HR diagram \citep{bagnulo03}. The  spectrum 
was obtained in 2002, September 18  with a resolving power of R\,=\,80\,000.

In order to determine the optimal parameters, we minimize the difference between
the observed and synthetic spectrum.  Thus we minimize
$$\chi^2 = \frac{1}{N} \sum \left(\frac{I_{\rm obs} - I_{\rm th}}{\delta I_{\rm
obs}}\right)^2$$
where $N$ is the total number of points, $I_{\rm obs}$ and $I_{\rm th}$ are the 
intensities of the observed and computed profiles, respectively, and $\delta I_{\rm obs}$ 
is the photon noise.  Synthetic spectra were generated in three  steps.   Firstly, 
we computed a model atmosphere using the ATLAS9 code.   The stellar spectrum was 
then synthesized using SYNTHE \citep{kur81}.  Finally, the spectrum was convolved 
with the instrumental and rotational profiles.

As starting values of T$_{\rm eff}$ and $\log g$, we used the values derived
from Str\"omgren photometry: $V = 6.075\pm0.007$, $b-y = 0.164\pm0.004$, 
$m_1 = 0.249\pm0.015$, $c_1 = 0.847\pm0.007$, $\beta = 2.813\pm0.005$ \citep{hauck98}.  
Using the algorithm in \citet{moon85} to obtain the reddening, we obtain $V_0 = 5.972$, $(b-y)_0 =
0.140$, $m_0 = 0.256$, $c_0 = 0.842$.  Note, however, that because of increased 
line blanketing in HD\,27411 relative to normal stars, it is not clear if the
de-reddening procedure can be applied (see below).  Using these values, the 
calibration of \citet{md85} leads to T$_{\rm eff}$\,=\,7820\,K and 
$\log g$\,=\,4.12, while the calibration of \citet{balona94} gives 
T$_{\rm eff}$\,=\,7760\,K, $\log g$\,=\,4.11.  Again, because of increased 
line  blanketing, these must be taken merely as provisional values.  We adopt 
T$_{\rm eff}$\,=\,7800\,K, $\log g$\,=\,4.12.  

To decrease the number of parameters, we computed the $v_e \sin i$ of 
HD\,27411 by matching synthetic line profiles from SYNTHE to a number of 
metallic lines.  The Mg{\sc i} triplet at $\lambda \lambda$5167-5183 {\AA} 
is particularly useful for this purpose.  The best fit was obtained with
$v_e \sin i$\,=\,20.5\,$\pm$\,0.5~km~s$^{-1}$.  This value is in good
agreement with $v_e \sin i$\,=\,20.4\,$\pm$\,0.4~km~s$^{-1}$ by
\citet{diaz11}.  

To determine stellar parameters as consistently as possible with the actual 
structure of the  atmosphere, we performed the abundance analysis by 
the following iterative procedure: 

\begin{description}

\item{(i)} $T_{\rm eff}$ is estimated by computing the ATLAS9 model atmosphere
which gives the best match between the observed H$_{\delta}$, H$_{\gamma}$, 
H$_{\beta}$, and H$_{\alpha}$ line profiles and those computed with SYNTHE.  
$\log g$ is estimated by matching the observed and calculated profiles of the 
Mg{\sc i} triplet at $\lambda\lambda$~5167--5183 {\AA} which is extremely 
sensitive to gravity.  This leads to T$_{\rm eff}$\,=\,7600\,$\pm$\,150~K, 
$\log g$\,=\,4.0\,$\pm$\,0.1, and ODF\,=\,[+0.5]. 

\item{(ii)} The microturbulent velocity, $\xi$, is determined independently 
from three sets of spectral lines: 71 lines from Fe{\sc i}, 15 lines from  
Fe{\sc ii}, and 24 lines from Ni{\sc i}.  For this purpose we used all lines 
with equivalent width (EW) $>$ 10 m{\AA}.  In particular, $\xi$ is computed by 
requiring that the derived abundances do not depend on the measured equivalent widths. 
To convert equivalent widths to abundances we used the WIDTH9 code \citep{kur81}.  
Values of $\xi$, the abundance and  the radial velocity obtained for each ion are listed 
in Table~\ref{xi_vr}. The adopted microturbulence is in agreement with \citet{land09} 
(see Fig. 2 of their paper).  The effective temperature is in agreement with the 
equilibrium condition between Fe{\sc i} and Fe{\sc ii}, since the iron abundances 
derived separately from these two different ionization stages are in good agreement.

\item{(iii)} The projected rotational velocity is relatively high.  To
overcome line blending problems, we divided the spectrum into a number of 
sub-intervals $\approx$25 {\AA} wide.  For each interval we performed a 
separate synthesis analysis. We used the abundances of Fe and Ni given in 
Table~\ref{xi_vr} as starting values in this procedure. The atomic parameters 
adopted in this analysis are from \citet{kur95} with line lists subsequently 
updated by \citet{castelli04}. The adopted abundances, shown in the second 
column of Table~\ref{abund}, are weighted averages expressed in the usual 
form $\log N_{\rm el}/N_{\rm Tot}$.

\end{description}

\begin{figure}
\centering
\includegraphics[width=8.5cm,bb=18 340 592 718]{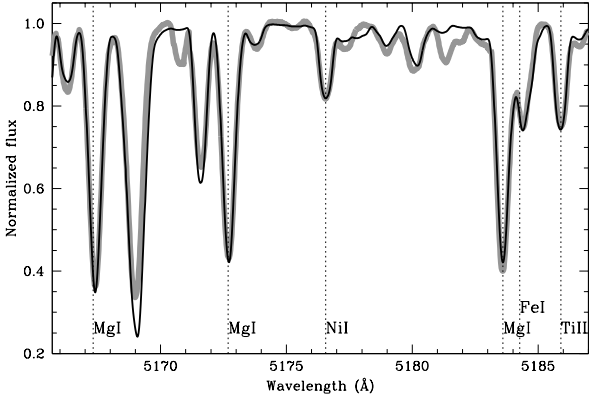}
\caption{Comparison between observed (crosses) and computed (solid red line) spectra in 
the region of the Mg{\sc i}b triplet, $\lambda \lambda$5167.321, 5172.684, and 5183.604. The stellar parameters
are T$_{\rm eff}$\,=\,7400 K, $\log g$\,=\,4.0, and $\log Mg/N_{\rm tot}$\,=\,$-$4.58\,$\pm$\,0.18.}
\label{MgIb}
\end{figure}

Values of T$_{\rm eff}$, $\log g$, $\xi$ and individual abundances estimated 
in this way were then used as initial guesses for starting another iterative 
procedure based on the ATLAS12 code.   The best fit was obtained after three 
iterations and led to the following parameters: T$_{\rm eff}$\,=\,7400 $\pm$ 150 K,
$\log g$\,=\,4.0 $\pm$ 0.1, $\xi$\,=\,4.2\,$\pm$\,0.3 km s$^{-1}$ and
$v_e \sin i$\,=\,20.5\,$\pm$\,0.5~km~s$^{-1}$.  The corresponding abundances 
are shown in the second column of Table~\ref{abund}.  \citet{ryab08}
analyzed HD\,27411 in a study of Ap stars and derived the following parameters: 
T$_{\rm eff}$\,=\,7650~K, $\log g$\,=\,4.0, $v_e \sin i$\,=\,18.5~km~s$^{-1}$, 
and $\xi$\,=\,2.5~km~s$^{-1}$. Considering the experimental errors, these values
are in agreement with ours.

The fits between the observed and synthetic Balmer lines are shown in Fig.~\ref{balmer}. 
The determination of surface gravity was constrained by using the Mg{\sc i} triplet 
at $\lambda\lambda$~5167--5183 {\AA}, as shown in Fig.~\ref{MgIb}. Errors in
T$_{\rm eff}$ and $\log g$ were estimated by the change in parameter values
which leads to an increase of $\chi^2$ by unity \citep{lampton76}.

\begin{figure}
\centering
\includegraphics[width=8.5cm]{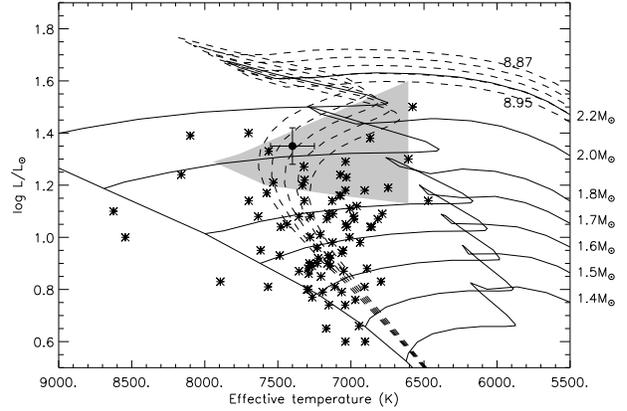}
\caption{Location of HD\,27411 in the HR diagram together with evolutionary
tracks and isochrones for $\log t$ ranging from 8.87 to 8.95 (step 0.02 and $t$ in yrs).
Asterisks are the pulsating AmFm stars taken from \citet{smalley11}. The gray area
is the approximate location of pulsating Am star models incorporating heavy-metal
diffusion \citep{turcotte00}.}
\label{HR}
\end{figure}

\begin{table}
 \centering
  \caption{Comparison among atmospheric parameters and abundances derived by ATLAS9 modeling and
           by ATLAS12 approach. The measurement without error is to be considered only as upper limit.
           In the last column we reported abundances derived with NLTE approach. All the abundances are
           expressed in the usual form $\log N_{\rm el}/N_{\rm Tot}$.}
  \begin{tabular}{lccc}
  \hline
  \hline
      & A9 & A12 & A12+SYNSPEC\\
 \hline
 T$_{\rm eff}$ & 7600\,$\pm$\,150  &  7400\,$\pm$\,150 & 7400\,$\pm$\,150\\
 $\log g$      & 4.0\,$\pm$\,0.1   &  4.0\,$\pm$\,0.1  & 4.0\,$\pm$\,0.1 \\
 \hline
 Li & $-$8.35\,$\pm$\,0.10 & $-$8.42\,$\pm$\,0.10 &  \\
 C  & $-$3.97\,$\pm$\,0.13 & $-$3.90\,$\pm$\,0.16 &  \\
 N  & $-$4.60              & $-$4.50              &  \\
 O  & $-$3.55\,$\pm$\,0.15 & $-$3.70\,$\pm$\,0.10 &  \\
 Na & $-$5.10\,$\pm$\,0.19 & $-$5.50\,$\pm$\,0.19 & $-$5.70\,$\pm$\,0.10 \\
 Mg & $-$4.46\,$\pm$\,0.13 & $-$4.58\,$\pm$\,0.18 &  \\
 Al & $-$5.11\,$\pm$\,0.17 & $-$5.05\,$\pm$\,0.19 & $-$5.44\,$\pm$\,0.10 \\
 Si & $-$4.21\,$\pm$\,0.16 & $-$4.19\,$\pm$\,0.10 & $-$4.49\,$\pm$\,0.12 \\
 S  & $-$4.54\,$\pm$\,0.14 & $-$4.52\,$\pm$\,0.18 & $-$4.75\,$\pm$\,0.10 \\
 K  & $-$6.85\,$\pm$\,0.10 & $-$6.70\,$\pm$\,0.10 &  \\
 Ca & $-$6.01\,$\pm$\,0.18 & $-$6.05\,$\pm$\,0.14 &  \\
 Sc & $-$9.23\,$\pm$\,0.10 & $-$9.15\,$\pm$\,0.17 &  \\
 Ti & $-$6.80\,$\pm$\,0.15 & $-$6.80\,$\pm$\,0.17 &  \\
 V  & $-$7.33\,$\pm$\,0.17 & $-$7.05\,$\pm$\,0.18 &  \\
 Cr & $-$5.72\,$\pm$\,0.13 & $-$5.70\,$\pm$\,0.15 & $-$5.95\,$\pm$\,0.10 \\
 Mn & $-$6.27\,$\pm$\,0.10 & $-$6.18\,$\pm$\,0.17 &  \\
 Fe & $-$4.07\,$\pm$\,0.12 & $-$4.19\,$\pm$\,0.13 &  \\
 Co & $-$6.07\,$\pm$\,0.11 & $-$6.19\,$\pm$\,0.17 &  \\
 Ni & $-$5.00\,$\pm$\,0.15 & $-$5.03\,$\pm$\,0.11 &  \\
 Cu & $-$6.93\,$\pm$\,0.16 & $-$6.94\,$\pm$\,0.11 &  \\
 Zn & $-$6.67\,$\pm$\,0.10 & $-$6.70\,$\pm$\,0.18 &  \\
 Y  & $-$8.77\,$\pm$\,0.10 & $-$8.84\,$\pm$\,0.18 &  \\
 Zr & $-$8.74\,$\pm$\,0.10 & $-$8.50\,$\pm$\,0.15 &  \\
 Ba & $-$9.01\,$\pm$\,0.18 & $-$9.15\,$\pm$\,0.15 &  \\
 La & $-$9.22\,$\pm$\,0.11 & $-$8.70\,$\pm$\,0.15 &  \\
 Ce & $-$8.83\,$\pm$\,0.15 & $-$8.55\,$\pm$\,0.15 &  \\
 Nd & $-$9.37\,$\pm$\,0.06 & $-$9.32\,$\pm$\,0.10 &  \\
 Sm & $-$9.53\,$\pm$\,0.06 & $-$9.47\,$\pm$\,0.09 &  \\
 Eu &$-$10.07\,$\pm$\,0.05 & $-$9.98\,$\pm$\,0.14 &  \\
\hline
\end{tabular}
\label{abund}
\end{table}

\section{Fundamental astrophysical quantities}
\label{age}

If we adopt $T_{\rm eff} = 7400 \pm 150$~K and $\log g = 4.00 \pm 0.10$ from
our spectroscopic analysis, we may use the relationships by \citet{torres10}
to derive $\log L/L_\odot = 0.99 \pm 0.12$.  These relate the mass and
radius of a star to the effective temperature and gravity through empirical
calibrations.  The greatest source of uncertainty is the surface gravity
determination.

The {\it Hipparcos} parallax for HD\,27411, $\pi$\,=\,11.13\,$\pm$\,0.38 
\citep{van07}, is useful in refining the location of the star in the HR diagram.
We show below that some caution is required in estimating reddening in Am/Fm
stars and that HD\,27411 is not significantly reddened.  We therefore adopt
$V_0 = 6.075$, which gives an absolute magnitude $M_V = 1.31 \pm 0.08$ where 
the error is derived from the error in the parallax. If we adopt the 
bolometric correction BC = 0.051 derived from \citet{balona94}, we have 
$M_{\rm bol} = 1.36 \pm 0.09$.  Using M$_{\rm bol,\odot}$\,=\,4.74 
\citep{drilling99}, we obtain $\log(L/L_\odot) = 1.35 \pm 0.07$.  From the 
luminosity and using $T_{\rm eff}$\,=\,7400~K, we obtain $R/R_\odot = 2.88 \pm 0.10$.
The surface gravity obtained by using the parameters derived from the parallax and 
assuming a mass of about $2.17 \pm 0.05~M_\odot$ is $\log g \approx 3.9 \pm 0.1$.
All the astrophysical quantities derived here are summarized in Table~\ref{phot}.

The location of the star in the HR diagram, together with some evolutionary 
tracks computed for non-solar metallicity Z\,=\,0.03 \citep{girardi00}, is shown in Fig.~\ref{HR}.  
Also shown are isochrones computed by \citet{marigo08} for the same Z and for five ages, i.e.
$\log t$\,=\,8.87, 8.89, 8.91, 8.93 and 8.95 ($t$ in years). The  non-solar metallicity follows 
from our abundance analysis, and Z=0.03 is the nearest metallicity in the models computed by 
\citet{girardi00} and \citet{marigo08}. The location of the star indicates a mass M\,$\approx$2\,M$_\odot$ 
and an age of $t \approx$\,810\,$\pm$\,40~Myrs. 

In Table~\ref{phot} we summarized all the astrophisical quantities for HD\,27411.

It is well known that the vast majority of Am stars are binaries.  There are 
few radial velocity measurements of HD\,27411 in the literature.  We could
only find three measurements in \citet{buscombe63}: +17, +34 and $-$53~km~s$^{-1}$.
{\bf From our spectrum, we derived a radial velocity RV\,=\,+40.6~$\pm$~0.3~km~s$^{-1}$, 
which is a weighted average of the single velocities derived by
converting the shift between observed and theoretical $\lambda_c$ 
from the species reported in Tablle~\ref{xi_vr}.} The scatter in these values does, 
indeed, suggest that HD\,27411 is a binary.  If so, then the companion is probably 
much fainter than the primary, otherwise one may expect to see some evidence in the
spectrum.

\begin{table}
\caption{Astrophysical quantities for HD\,27411.  The second column are
quantities derived from the parallax.  The third column are quantities
derived from Str\"{o}mgren photometry and the last column from
spectroscopy.}
\label{phot} 
\centering         
\begin{tabular}{lrrr}
\hline
\hline
Paramerter       &  Parallax         &  Str\"{o}mgren   & Spectroscopy    \\
\hline
$\pi$            & $11.13 \pm 0.38$  &                  &                 \\
$V_0$            & $6.075 \pm 0.007$ & $5.972 \pm 0.005$&                 \\
M$_V$            & $1.31  \pm 0.08$  &                  &                 \\
M$_{\rm bol}$    & $1.36  \pm 0.09$  &                  &                 \\
T$_{\rm eff}$    &                   & $7800 \pm 200$   & $7400 \pm 150$  \\
$\log(L/L_\odot)$& $1.35  \pm 0.07$  & $0.96 \pm 0.15$  & $0.99 \pm 0.12$ \\
$\log g$         & $3.9 \pm 0.1$     & $4.12 \pm 0.10$  & $4.00 \pm 0.10$ \\
R/R$_\odot$      & $2.88 \pm 0.10$   & $1.91 \pm 0.27$  & $1.90 \pm 0.26$ \\
$M/M_\odot$      & $2.17 \pm 0.05$   & $1.77 \pm 0.07$  & $1.66 \pm 0.07$ \\
\hline
\hline
\end{tabular}
\end{table}

\section{Abundances}

\begin{figure}
\centering
\includegraphics[width=8.5cm]{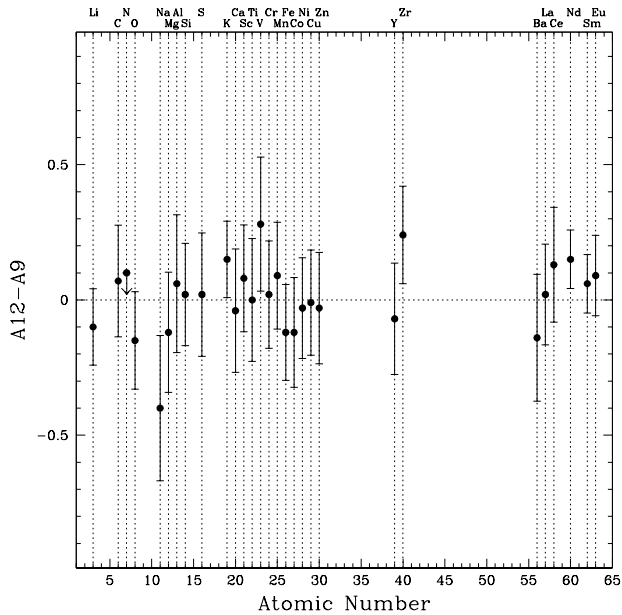}
\caption{Comparison between abundances computed using ATLAS9 and ATLAS12 
model atmospheres. The differences in abundance given by the two models are
shown as a function of atomic number.}
\label{A12_A9}
\end{figure}

\begin{figure}
\centering
\includegraphics[width=8.5cm]{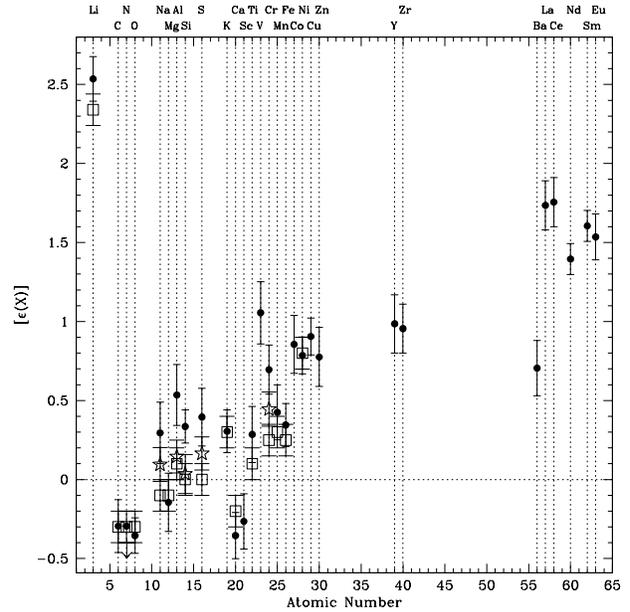}
\caption{Abundances in HD\,27411 as a function of atomic number. Filled 
circles represent our abundances.  Open boxes (red) are abundances 
predicted by the diffusion model of \citet{richer00} for 670~Myr.
Starred symbols represent the elements for which NLTE abundances
have been calculated.}
\label{pattern}
\end{figure}

In Fig.~\ref{A12_A9} we compare the LTE abundances derived from
model atmospheres computed using ATLAS9 and ATLAS12. It is evident that there 
is good agreement in the abundances derived with the two different codes. 
There are some very small differences for Na, V, Zr and Nd, but these are
only 0.1~dex or less.  Thus we may use the faster ATLAS9 code with
confidence.  In Fig.~\ref{pattern} we show the abundances relative to solar 
standard abundances \citep{grevesse10}.  The chemical pattern displayed here
is typical of that observed in Am stars, i.e. an underabundance of C, N, O, 
Ca and Sc and a general increasing overabundance for  heavy elements.

The atmospheric abundance of Li is interesting in the context of
diffusion.  \citet{burkhart91} and \citet{burkhart05} find that, in 
general, the Li abundance in Am stars is close to the cosmic
value ($\log N_{\rm Li}/N_{\rm Tot} \approx -9.04$~dex), although some Am/Fm stars appear
to have an underabundance of Li.  Normal A-type stars in the range
$7000 < T_{\rm eff} < 8500$~K appear to have a higher Li abundance, i.e. 
$\log N_{\rm Li}/N_{\rm Tot} \approx -8.64$~dex, \citep{burkhart95}.  To determine the 
Li abundance, we used the Li{\sc i} $\lambda$6707 {\AA} line, taking into 
account the  hyperfine structure \citep{andersen84}. The abundance that 
gives the best fit is $\log N_{\rm Li}/N_{\rm Tot}$\,=\,$-$8.42$\pm$0.10, 
which is closer to the abundance in normal A stars and agrees with the
average Li abundance of three cluster Am/Fm stars observed by 
\citet{fossati07}.

\citet{richer00} predict abundances as a function of stellar age and
effective temperature using their models of diffusion.   Fig.~14 of their
paper allows us to estimate the predicted abundances for a star with a 
given T$_{\rm eff}$ up to a maximum age of $\approx$\,670~Myrs.  We find
that age of HD\,27411 to be about $810 \pm 40$~Myr, which is considerably
older than the maximum age of models in \citet{richer00}, but we will assume
that models of 670~Myr still give a fair approximation of the abundances.  
For T$_{\rm eff}$\,=\,7400~K, which is our best estimate for HD\,27411, 
the models by \citet{richer00} predicts underabundances ranging from -0.3~dex 
to -0.1~dex for C, N, O, Na, Mg, K, and Ca.  For Si and S the abundances 
are normal, while overabundances of about 0.1--0.8~dex are found for for 
Li, Al, Ti, Cr, Mn, Fe, and Ni.  Inspection of their figure reveals that 
for Na, Mg, Al, Si, S, Ca, Ti, Cr, Mn, and Fe, the abundance anomaly is 
approximately constant with age and depends only on the turbulence.   The 
abundances of Li, C, N, and O  vary with age.

Fig.~\ref{pattern} shows the abundances of the elements at 670~Myr predicted 
by \citet{richer00} compared with our abundances. There is indeed good 
agreement for Li, C, N, O, Mg, K, Ca, Ti, Mn, Fe, and Ni, but abundances of 
Na, Al, Si, S, and Cr are somewhat discrepant. Similar discrepancies for Na and
S were found by \citet{fossati07} for the Am star HD\,73730.  From the NLTE 
analysis of the S abundance by \citet{kamp01}, \citet{fossati07} concluded that
NLTE effects  should be taken into account to determine whether this resolves 
the discrepancy with diffusion predictions. 

Following this idea, we performed a NLTE analysis on Na, Al, Si, S, and Cr, 
to derive their abundances.  We used the same technique of matching predicted 
and observed line profiles, but in this case the NLTE line profiles were 
computed with version 43 of SYNSPEC \citep{hubeny00}. This code reads the same
input model atmosphere previously computed using ATLAS12 and solves the 
radiative transfer equation, wavelength by wavelength in a specified spectral 
range. SYNSPEC also reads the same Kurucz list of lines that we used for 
determining metal abundances.  SYNSPEC allows one to compute the line profiles 
by using an approximate NLTE treatment, even for LTE models. This is done by 
means of second-order escape probability theory (for details see 
\citet{hubeny86}). The results of these calculations are shown in 
Table~\ref{abund}. All the NLTE abundances are lower than the LTE abundances 
by factors ranging from 0.23~dex (S) to 0.39~dex (Al). As can be seen from 
Fig.~\ref{pattern}, the NLTE calculations bring the observations closer to the
diffusion predictions by \citet{richer00}.  In fact, there is no longer any 
discrepancy in abundances within the observational errors.

\section{Effect of line blanketing on the photometry}

In order to investigate the effect of line blanketing on the Str\"{o}mgren
colour indices, we used the method of synthetic photometry.  For this
purpose we computed the spectrum of a star at different effective
temperatures with abundances shown in Table\,\ref{abund}. 
We used the abundances given by the ATLAS12 models modified by NLTE where
necessary.  The spectra were calculated using {\tt SPECTRUM}, version 
2.76e\footnote{{\tt www1.appstate.edu/dept/physics/spectrum/spectrum.html}} 
\citep{gray94}.  Synthetic spectra with normal and peculiar abundances
were calculated for $6000 \le T_{\rm eff} < 10\,000$~K and $\log g =
4.00$.  In all cases the microturbulence velocity was set to $\xi=4$~km~s$^{-1}$.   
These spectra were convolved with standard $uvby\beta$ transmission functions 
to calculate synthetic Str\"{o}mgren indices. 

It should be noted that not all Am/Fm stars will have the same abundance
anomalies as HD\,27411.  Hence the results described here are only
indicative of what might be typical in Am/Fm stars.  Individual Am/Fm stars
will have different abundances and different line blanketing.

\begin{figure}
\centering
\includegraphics[width=8.5cm]{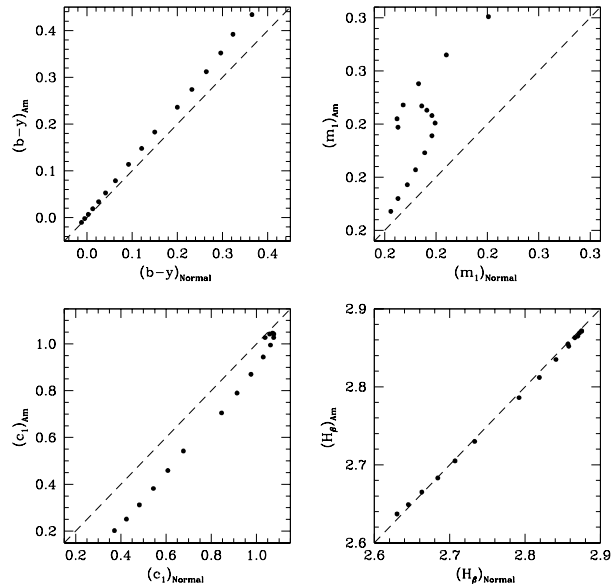}
\caption{Relation between synthetic Str\"{o}mgren indices for normal stars
(horizontal axis) and for Am/Fm stars (vertical axis).  The straight line
represents equality between the indices.}
\label{colcol}
\end{figure}

In computing these synthetic Str\"{o}mgren indices, it is necessary to
identify a particular model with a real star in order to determine the zero
points.  We chose a model of Vega ($b-y = 0.003, m_1 = 0.157, c_1 = 1.088, 
\beta = 2.903$) for this purpose.  Comparison of the synthetic $b-y$ as a 
function of $\beta$ with the standard relations of \citet{crawford75} shows 
that the $\beta$ zero point required a further correction of -0.04.  The 
synthetic colours for normal and Am/Fm stars are listed in
Table\,\ref{strom}.   Comparison of indices for models with standard solar
abundance and the abundances of Table\,\ref{abund} are shown in
Fig.\,\ref{colcol}.

\begin{table}
 \centering
  \caption{Synthetic Str\"{o}mgren indices for normal stars and Am/Fm stars
with the abundances of Table\,\ref{abund}.}
\begin{tabular}{rrrrr}
\hline
\hline
$T_{\rm eff}$ & $b-y$ & $m_1$ & $c_1$ & $\beta$ \\
 \hline
  \\
Normal:\\
  6000 &  0.365 & 0.251 & 0.372 & 2.630  \\
  6250 &  0.323 & 0.210 & 0.425 & 2.645  \\
  6500 &  0.296 & 0.183 & 0.482 & 2.663  \\
  6750 &  0.264 & 0.168 & 0.544 & 2.684  \\
  7000 &  0.232 & 0.162 & 0.608 & 2.707  \\
  7250 &  0.200 & 0.163 & 0.677 & 2.733  \\
  7500 &  0.150 & 0.186 & 0.846 & 2.792  \\
  7750 &  0.121 & 0.191 & 0.914 & 2.819  \\
  8000 &  0.092 & 0.196 & 0.976 & 2.841  \\
  8250 &  0.063 & 0.199 & 1.030 & 2.858  \\
  8500 &  0.041 & 0.196 & 1.062 & 2.870  \\
  8750 &  0.026 & 0.189 & 1.076 & 2.875  \\
  9000 &  0.013 & 0.180 & 1.077 & 2.875  \\
  9250 &  0.003 & 0.172 & 1.071 & 2.872  \\
  9500 & -0.005 & 0.163 & 1.058 & 2.866  \\
  9750 & -0.012 & 0.156 & 1.038 & 2.857  \\
\\
Am/Fm:\\
  6000 &  0.434 & 0.351 & 0.202 & 2.637  \\
  6250 &  0.392 & 0.315 & 0.251 & 2.649  \\
  6500 &  0.352 & 0.288 & 0.312 & 2.665  \\
  6750 &  0.312 & 0.268 & 0.382 & 2.683  \\
  7000 &  0.274 & 0.255 & 0.459 & 2.705  \\
  7250 &  0.236 & 0.247 & 0.542 & 2.730  \\
  7500 &  0.183 & 0.267 & 0.705 & 2.786  \\
  7750 &  0.148 & 0.263 & 0.790 & 2.812  \\
  8000 &  0.114 & 0.258 & 0.870 & 2.835  \\
  8250 &  0.079 & 0.251 & 0.944 & 2.852  \\
  8500 &  0.053 & 0.239 & 0.995 & 2.865  \\
  8750 &  0.034 & 0.223 & 1.027 & 2.871  \\
  9000 &  0.019 & 0.207 & 1.042 & 2.872  \\
  9250 &  0.007 & 0.193 & 1.046 & 2.869  \\
  9500 & -0.002 & 0.180 & 1.042 & 2.863  \\
  9750 & -0.010 & 0.168 & 1.027 & 2.855  \\
\hline
\end{tabular}
\label{strom}
\end{table}

\begin{figure}
\centering
\includegraphics[width=8.5cm]{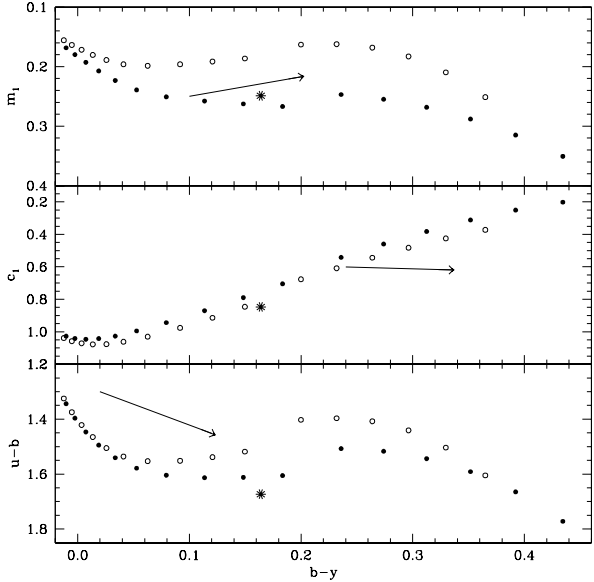}
\caption{Synthetic colour--colour diagrams for stars with normal abundance 
(open circles) and with abundances of Table\,\ref{abund} (filled circles).
The asterisk shows the observed location of HD\,27411 and the arrows are the
reddening lines from \citet{crawford76}.}
\label{red}
\end{figure}

Various colour--colour diagrams derived from the synthetic colours are shown
in Fig.\,\ref{red} together with reddening lines.  The reddening lines are
from \citet{crawford76}:\\
\\
$\frac{E(u-b)}{E(b-y)} = 1.53$, $\frac{E(c_1)}{E(b-y)} = 0.19$,
$\frac{E(m_1)}{E(b-y)} = -0.33.$\\
\\
Also shown are the locations of HD\,27411.  It is
clear that the star lies nicely on the synthetic relations given by the
enhanced abundances and that the star is unreddened or only very slightly
reddened.  It is also clear that in determining the reddenings of Am/Fm
stars it is wise to avoid using $m_1$ and $u-b$.  The $c_1/b-y$ relations
for Am/Fm abundances is almost the same as for normal abundances and it is
preferable to use this diagram to deduce the reddening correction. 
Assuming that HD\,27411 is unreddened and matching the observed indices with
those for modified abundances in Table\,\ref{strom} gives $T_{\rm eff} =
7600$~K.

The $\beta$ index is correlated well with effective temperature for $T_{\rm
eff} < 8000$~K and is practically unaffected by line blanketing.  It is
weakly sensitive to surface gravity, particularly for the hotter stars. 
$b-y$ is correlated with effective temperature but is affected by blanketing
for the cooler stars. As can be seen from Fig\,\ref{colcol}, the $m_1$ index 
is severely affected by blanketing. The $c_1$ index for Am/Fm stars is nearly 
always smaller than for normal stars.  This index measures the strength of the
Balmer discontinuity (and hence the surface gravity), but it is clearly not 
entirely free of blanketing effects.

In estimating the absolute magnitudes, $M_V$, of F stars, \citet{crawford75} 
defines, first of all, a relationship for stars on the ZAMS as a function of
$b-y$, $M_V = M_V(ZAMS, b-y).$   The absolute magnitude for a star above the 
ZAMS is calculated from the value of $\delta c_1 = c_1 - c_1(ZAMS)$, i.e.
the difference between the measured $c_1$ and the value of $c_1$ on the ZAMS
at the given $b-y$.  Since there are significant line blanketing effects on
$c_1$ for Am/Fm stars, their absolute magnitudes derived in this way are
probably not free of systematic errors.  We derived $M_V$ using the data of
Table\,\ref{strom} for stars with solar abundance and with the Am abundance
using the calibration of \citet{crawford75}.  We find that on average the Am
stars are estimated to be about 1.2 magnitudes fainter than normal stars of
the same effective temperature and gravity.  This is due to the
systematically lower values of $c_1$ in the Am star models. 
Table\,\ref{groundspec} lists Am stars which have good parallaxes.  The
general trend of lower luminosities derived from Str\"{o}mgren photometry is
apparent.

From this exercise we conclude that although the effective temperatures of 
Am/Fm stars derived from Str\"{o}mgren photometry are probably reliable, the 
absolute magnitudes may be systematically too faint.  {\bf For example, if 
we apply the \citet{crawford75} calibration to HD\,27411, assuming no reddening, 
we obtain $M_V = 2.34$ or $\log L/L_\odot = 0.96$ (using BC = 0.035 derived 
from \citet{balona94}), whereas the most reliable estimate (Hipparcos parallax) 
gives $\log L/L_\odot = 1.39$.} This effect can be seen in Fig\,\ref{HR} where 
many of the Am stars are below the line defining the ZAMS.  The luminosities 
of these stars were derived using the standard calibration and hence have been 
under-estimated. 

\begin{table} 
\caption{Am stars which have very good trigonometric parallaxes.  The 
luminosities calculated from the parallaxes and from Str\"{o}mgren
photometry are shown.}
\label{groundspec}
\begin{center} 
\begin{tabular}{rl@{ }rr@{ }r@{ }r}  
\hline 
\hline 
    &                &                        &          & Parallax & Str\"{o}mgren \\
 HD & Class & $\log\rm{T}_{\rm eff}$ & $\log g$ & $\log \frac{L}{L_\odot}$ & $\log \frac{L}{L_\odot}$ \\
\hline                            
\noalign{\smallskip}              
   27411 & A3m        &  3.869 & 4.12  & 1.35 &  0.96  \\   
   27628 & kA5hF0mF2  &  3.864 & 4.00  & 1.04 &  0.82  \\
   71297 & A5III-IV   &  3.900 & 4.19  & 1.14 &  0.82  \\
  104513 & A7m        &  3.879 & 4.24  & 0.88 &  0.87  \\
  204188 & kA6hA9mF0  &  3.898 & 4.34  & 0.89 &  0.74  \\
\noalign{\smallskip}  
\hline 
\end{tabular}  
\end{center} 
\end{table}

\section{Inferences from pulsations in Am/Fm stars}

The diffusion models of \citet{richer00} are the best models presently
available for Am/Fm stars.  The models seem to be able to predict the
abundances in these stars rather well, but we need to bear in mind that
this is achieved because of adjustable free parameters to describe the 
turbulence.  We still do not know if the mechanism competing with diffusion
is turbulence, mass loss or some other factor.  What we do know is that the
current description of Am stars is in trouble because it fails to account
for the wide distribution of pulsating Am stars in the $\delta$~Sct
instability strip \citep{balona11,smalley11}. 

One question that is of interest is the fraction of Am stars that pulsate.
To answer this question we have to define what we mean by ``non-pulsating''.  
Clearly, a star could be pulsating but with amplitudes too small to be visible 
from the ground.  \citet{balona11b} discussed this issue in the context of 
{\it Kepler} observations which, of course, allow pulsations to be detected at 
the micromagnitude level.  They deduced that the fraction of pulsating stars in 
the $\delta$~Sct instability strip is surprisingly low.  There is clearly
some damping mechanism which is currently not understood.  The fraction of
$\delta$~Sct stars in the instability strip varies with effective
temperature, but does not exceed about 50 percent.

We can answer this question for Am stars only in part because we do not have
a sufficient number of Am stars observed at the micromagnitude level.  In
order to compare these ground-based observations with the extensive {\it
Kepler} observations of $\delta$~Sct stars, we need to degrade the {\it
Kepler} data by considering as pulsating only those $\delta$~Sct stars with 
amplitudes over 1.5~mmag.  We chose this minimum amplitude as roughly
representative of the detection limit in the catalogue of pulsating Am stars
in \citet{smalley11}.  The percentage of {\it Kepler} $\delta$~Sct stars
with this minimum amplitude relative to all stars in a particular temperature 
range is shown in Table\,\ref{dsct}.

\begin{table} 
\caption{The numbers of all stars in the given effective temperature range,
$N$(All), and the number of $\delta$~Sct stars in that range,
$N$($\delta$~Sct), and the corresponding percentage is given.  The first
block refers to {\it Kepler} $\delta$~Scuti stars with maximum amplitude 
exceeding 1.5~mmag.  The second block refers to all Am stars and the pulsating 
Am stars in the given temperature range.}
\label{dsct}
\begin{center} 
\begin{tabular}{lrrr}
  \hline
  \hline
$T_{\rm eff}$ & $N$(All) & $N$($\delta$~Sct) & Percent \\
\hline
{\it Kepler}:\\
 5500 -  6500 & 1509 &  33 &  2.19 \\
 6500 -  7000 & 3842 & 174 &  4.53 \\
 7000 -  7500 & 1412 & 263 & 18.63 \\
 7500 -  8000 &  811 & 172 & 21.21 \\
 8000 -  8500 &  512 &  51 &  9.96 \\
 8500 -  9000 &  297 &  11 &  3.70 \\
 9000 - 10000 &  263 &   1 &  0.38 \\
\\
Am stars:\\
 5500 -  6500 &   16 &   2 & 12.50 \\
 6500 -  7000 &   75 &  21 & 28.00 \\
 7000 -  7500 &  307 &  52 & 16.94 \\
 7500 -  8000 &  489 &   9 &  1.84 \\
 8000 -  8500 &  185 &   1 &  0.54 \\
 8500 -  9000 &   32 &   2 &  6.25 \\
 9000 - 10000 &    7 &   0 &  0.00 \\
\hline 
\end{tabular}  
\end{center} 
\end{table} 

\begin{figure}
\includegraphics[width=8.5cm,bb=18 340 592 718]{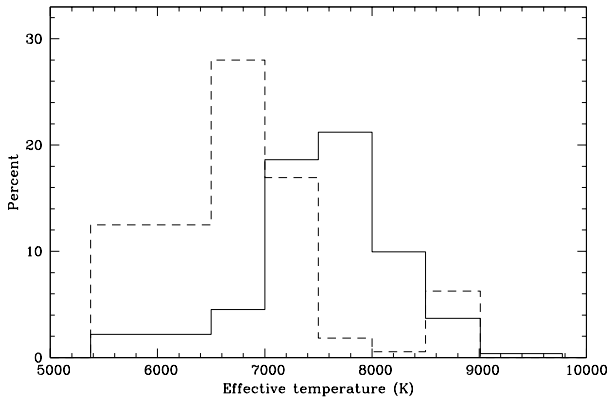}
\caption{The percentage of $\delta$~Scuti stars relative to all stars in the
given effective temperature range is shown as the solid line histogram.  The
dashed line histogram is the percentage of pulsating Am stars relative to
all Am stars in the given temperature range.}
\label{distr}
\end{figure}

We can now compare this distribution of ordinary $\delta$~Sct stars with
the distribution of pulsating Am stars in \citet{smalley11}.  We used 
the catalogue of \citet{renson09} and estimated the effective temperatures
of the Am stars using the \citet{balona94} calibrations.  Results are shown 
in Table\,\ref{dsct} and Fig.\,\ref{distr}.  It is evident that the pulsating 
Am stars are cooler than normal $\delta$~Sct stars, a fact already mentioned 
by \citet{smalley11}.  This conclusion still holds even if the amplitude
threshold in {\it Kepler} data is lowered to a few micromagnitudes.

From this comparison, we may deduce that there is certainly a tendency for
pulsating Am stars to be confined more towards the red edge, but that this
effect is far smaller than predicted by \citet{turcotte00}.  In fact,
Table\,\ref{dsct} shows that the percentage of pulsating Am stars among the
Am stars is about the same as the percentage of $\delta$~Scuti stars in the
instability strip.  This tells us that driving in the He{\sc ii} ionization
zone is practically unaffected.  We may therefore conclude that there is
no significant reduction of He in the ionization zone, contrary to the
prediction of current diffusion models.

\section{Conclusions}

We analyzed the spectrum of the Am star HD\,27411 from the UVES POP archive. 
Our aim was to investigate if the ATLAS12 model atmosphere code provides more 
reliable results than the ATLAS9 code for chemically peculiar stars. 
We found that there is very little difference in the abundances derived from ATLAS9
and from ATLAS12.   Since ATLAS12 demands considerably greater resources, it
seems safe to use ATLAS9, at least for moderate metallic enhancement.
We find that the derived abundances in HD\,27411 are in good agreement
with the predictions of diffusion models by \citet{richer00}.  There were 
discrepancies for Na, Al, Si, S, and Cr, but these are resolved by using
NLTE model atmospheres.

We investigated the reliability of effective temperatures and luminosities
of Am/Fm stars determined by Str\"{o}mgren photometry by synthesizing
spectra having the abundances of HD\,27411 for a range of effective
temperatures.  The resulting synthetic colours indicate that effective
temperatures can be reliably determined from photometry, but owing to line
blanketing in the $c_1$ passband, the resulting surface gravities are
systematically to high, leading to lower luminosities.  This result appears
to be verified by comparing luminosities of Am/Fm stars obtained from their
parallaxes and from photometry.

{\bf Determination of reliable luminosities for Am stars remains a difficult
problem.  At this stage, parallaxes offer the best results, but this can
only be done for very few stars.  As we have seen, luminosities obtained
from Str\"{o}mgren photometry are subject to a systematic bias which depends
on the overabundances of metals.  The error in the surface gravity from 
high-resolution spectroscopy is typically 0.1 in $\log g$.  For A--F main 
sequence and giants, this translates into an error of about 0.12 in 
$\log L/L_\odot$ when using the calibration of \citet{torres10}.  For 
HD\,27411, for example, we derive $\log g = 4.0 \pm 0.1$ from spectroscopy, 
whereas the value derived from the parallax is $\log g = 3.8 \pm 0.1$ (Table\,2),
Although the two values only differ by two standard deviations, this is
enough to cause a difference of 0.36 in $\log L/L_\odot$.  Although
spectroscopic determinations of luminosities may lead to quite large errors
in the luminosity, they should at least not be biased.}

By far the most serious problem confronting the diffusion model is that
there seems to be no appreciable settling of He in the He{\sc ii} ionization
zone, as predicted by the models. This is demonstrated by the fact that
pulsating Am/Fm stars occur throughout the $\delta$~Scuti instability strip,
though they tend to be cooler than normal $\delta$~Sct stars.  In fact, the 
relative proportions of pulsating Am stars to non-pulsating Am stars is no 
different from the proportion of $\delta$~Sct stars to constant
stars in the $\delta$~Sct instability strip.   There is clearly a need to
revise current ideas of diffusion to explain the Am phenomenon.

\section*{Acknowledgments}

LAB wishes to thank the National Research Foundation and the South African
Astronomical Observatory for financial assistance.

\label{lastpage}

\end{document}